\documentclass[review,number,sort&compress,fleqn]{elsarticle}

\usepackage{amsthm,amsmath}
\usepackage{caption}
\usepackage{lineno,hyperref}
\modulolinenumbers[5]

\usepackage{siunitx}
\usepackage{cases}
\usepackage{wasysym} 
\usepackage{placeins}
\usepackage{xcolor,ulem}

\journal{Nuclear Instruments and Methods in Physics Research Section A}

\begin{document}

\begin{frontmatter}

\title{Measurement of the neutron beam profile of the Back-n white neutron facility at CSNS with a Micromegas detector}

\author[keylab-ustc,ustc-dmp] {Binbin~Qi\fnref{cofirst}}
\author[ihep,dnsc] {Yang~Li\fnref{cofirst}}
\author[keylab-ustc,ustc-dmp] {Danyang~Zhu}
\author[keylab-ustc,ustc-dmp] {Zhiyong~Zhang\corref{cor}}
\ead{zhzhy@ustc.edu.cn}
\author[keylab-ihep,ihep,dnsc] {Ruirui~Fan\corref{cor}}
\ead{fanrr@ihep.ac.cn}
\author[keylab-ustc,ustc-dmp] {Jiang~Pan}
\author[keylab-ustc,ustc-dmp] {Jianxin~Feng}
\author[keylab-ustc,ustc-dmp] {Chengming~Liu}
\author[keylab-ustc,ustc-dmp] {Changqing~Feng}
\author[keylab-ustc,ustc-dmp] {Jianbei~Liu}
\author[keylab-ustc,ustc-dmp] {Ming~Shao}
\author[keylab-ustc,ustc-dmp] {Yi~Zhou}
\author[ihep,dnsc] {Yanfeng~Wang}
\author[ihep,dnsc] {Han~Yi}
\author[keylab-ustc,ustc-dmp]{Qi~An}
\author[pku]{Huaiyong~Bai}
\author[401]{Jie~Bao}
\author[keylab-ustc,ustc-dmp]{Ping~Cao}
\author[caep]{Qiping~Chen}
\author[ihep,dnsc]{Yonghao~Chen}
\author[usc]{Pinjing~Cheng}
\author[pku]{Zengqi~Cui}
\author[keylab-ihep,ihep]{Minhao~Gu}
\author[ihep,dnsc]{Fengqin~Guo}
\author[21inst]{Changcai~Han}
\author[caep]{Zijie~Han}
\author[401]{Guozhu~He}
\author[ihep,dnsc]{Yongcheng~He}
\author[usc]{Yuefeng~He}
\author[401]{Hanxiong~Huang}
\author[ihep,dnsc]{Weiling~Huang}
\author[keylab-ustc,ustc-dmp]{Xiru~Huang}
\author[keylab-ihep,ihep]{Xiaolu~Ji}
\author[keylab-ustc,ustc-deap]{Xuyang~Ji}
\author[pku]{Haoyu~Jiang}
\author[ihep,dnsc]{Wei~Jiang}
\author[ihep,dnsc]{Hantao~Jing}
\author[ihep,dnsc]{Ling~Kang}
\author[ihep,dnsc]{Mingtao~Kang}
\author[ihep,dnsc]{Bo~Li}
\author[ihep,dnsc]{Lun~Li}
\author[ihep,dnsc] {Qiang~Li}
\author[ihep,dnsc]{Xiao~Li}
\author[keylab-ihep,ihep]{Yang~Li}
\author[caep]{Rong~Liu}
\author[keylab-ustc,ustc-dmp]{Shubin~Liu}
\author[caep]{Xingyan~Liu}
\author[401]{Guangyuan~Luan}
\author[ihep,dnsc]{Yinglin~Ma}
\author[ihep,dnsc]{Changjun~Ning}
\author[401]{Jie~Ren}
\author[401]{Xichao~Ruan}
\author[21inst]{Zhaohui~Song}
\author[ihep,dnsc]{Hong~Sun}
\author[ihep,dnsc]{Xiaoyang~Sun}
\author[keylab-ihep,ihep,dnsc]{Zhijia~Sun}
\author[ihep,dnsc]{Zhixin~Tan}
\author[401]{Hongqing~Tang}
\author[ihep,dnsc]{Jingyu~Tang}
\author[ihep,dnsc]{Pengcheng~Wang}
\author[401]{Qi~Wang}
\author[buaa]{Taofeng~Wang}
\author[401]{Zhaohui~Wang}
\author[ihep,dnsc]{Zheng~Wang}
\author[caep]{Jie~Wen}
\author[caep]{Zhongwei~Wen}
\author[ihep,dnsc]{Qingbiao~Wu}
\author[401]{Xiaoguang~Wu}
\author[ihep,dnsc]{Xuan~Wu}
\author[keylab-ustc,ustc-deap]{Likun~Xie}
\author[caep]{Yiwei~Yang}
\author[ihep,dnsc]{Li~Yu}
\author[keylab-ustc,ustc-dmp]{Tao~Yu}
\author[ihep,dnsc]{Yongji~Yu}
\author[pku]{Guohui~Zhang}
\author[ihep,dnsc]{Jing~Zhang}
\author[ihep,dnsc]{Linhao~Zhang}
\author[keylab-ustc,ihep,dnsc]{Liying~Zhang}
\author[xju]{Qingming~Zhang}
\author[401]{Qiwei~Zhang}
\author[21inst]{Xianpeng~Zhang}
\author[ihep,dnsc]{Yuliang~Zhang}
\author[xju]{Yingtan~Zhao}
\author[ihep,dnsc]{Liang~Zhou}
\author[401]{Zuying~Zhou}
\author[keylab-ihep,ihep]{Kejun~Zhu}
\author[ihep,dnsc]{Peng~Zhu}

\cortext[cor]{Corresponding author.}
\fntext[cofirst]{Co-first author.  These authors contributed equally to this work.}
\address[keylab-ustc]{State Key Laboratory of Particle Detection and Electronics, University of Science and Technology of China, Hefei 230026, China}
\address[ustc-dmp]{Department of Modern Physics, University of Science and Technology of China, Hefei 230026, China}
\address[ihep]{Institute of High Energy Physics, Chinese Academy of Sciences, Beijing 100049, China}
\address[dnsc]{Spallation Neutron Source Science Center, Dongguan 523803, China}
\address[keylab-ihep]{State Key Laboratory of Particle Detection and Electronics, Institute of High Energy Physics, Chinese Academy of Sciences, Beijing 100049, China}
\address[pku]{State Key Laboratory of Nuclear Physics and Technology, School of Physics, Peking University, Beijing 100871, China}
\address[401]{Key Laboratory of Nuclear Data, China Institute of Atomic Energy, Beijing 102413, China}
\address[caep]{Institute of Nuclear Physics and Chemistry, China Academy of Engineering Physics, Mianyang 621900, China}
\address[usc]{University of South China, Hengyang 421001, China}
\address[21inst]{Northwest Institute of Nuclear Technology, Xi’an 710024, China}
\address[ustc-deap]{Department of Engineering and Applied Physics, University of Science and Technology of China, Hefei 230026, China}
\address[buaa]{School of Physics, Beihang University 100083, China}
\address[xju]{Xi’an Jiaotong University, Xi’an 710049, China}

\begin{abstract}
The Back-n white neutron beam line, which uses back-streaming white neutrons from the spallation target of the China Spallation Neutron Source, is used for nuclear data measurements.  A Micromegas-based neutron detector with two variants was specially developed to measure the beam spot distribution for this beam line.  In this article, the design, fabrication, and characterization of the detector are described.  The results of the detector performance tests are presented, which include the relative electron transparency, the gain and the gain uniformity, and the neutron beam profile reconstruction capability.  The result of the first measurement of the Back-n neutron beam spot distribution is also presented.
\end{abstract}


\end{frontmatter}


\section{Introduction}
The neutron beam line using back-streaming white neutrons (Back-n)~\cite{bib:HTJ2010, bib:JYT2010, bib:An2017} at China Spallation Neutron Source (CSNS)~\cite{bib:Chen2016, bib:Wang2009} was built and commissioned in 2018.  This beam line is called the \lq\lq Back-n white neutron beam line\rq\rq\ or simply \lq\lq Back-n\rq\rq.  Back-n delivers an intense neutron flux up to \SI{\sim e7}{\cm^{-2}.\s^{-1}} at a distance of \SI{55}{\m} from the spallation target at \SI{100}{\kW} proton beam power, and the neutrons span many orders of magnitude in energy, from \SI{0.5}{\eV} to several hundreds of MeV.

Equipped with several dedicated spectrometers, Back-n is a facility designed for nuclear data measurements and has been providing critical data for various research activities such as the design of nuclear energy production facilities, nuclear astrophysics research, and the improvement of the current nuclear databases which have very wide applications.  Measurements of the neutron beam properties, e.g. the spot distribution, the intensity, and the energy spectrum of the neutron beam are the basic requirement for this kind of facilities.  Thanks to the good two-dimensional (2D) spatial resolution and fast timing capability of the Micromegas (micro-mesh gaseous structure) detector~\cite{bib:GIOMATARIS1996, bib:PANCIN2004}, it can be used for a quasi-online measurement of the neutron beam profile~\cite{bib:Diakaki2018} at the Back-n white neutron facility.

Micromegas detectors (MMDs) have been widely used in nuclear and particle physics including rare-event searches~\cite{bib:Irastorza2016, bib:Irastorza2016e} and neutron detection~\cite{bib:Bellont2013}.  In this article, we present the development of an MMD with 2D spatial resolution capability for measuring the neutron beam spot distribution.  For a neutron to be detected, it needs to be converted into charged particles using nuclear reactions.  A thin layer of ${}^{10}\mathrm{B}$ or ${}^{6}\mathrm{Li}$ deposited on a thin aluminum foil is used as the neutron converter.  The current setup of the MMD, used as a beam profiler at Back-n, sums up the neutron events of the entire energy range, with higher weights on low-energy neutrons due to higher cross sections ${}^{10}\mathrm{B} (n,\,\alpha) {}^{7}\mathrm{Li}$ and ${}^{6}\mathrm{Li} (n,\,t) {}^{4}\mathrm{He}$.  However, it is planned, as part of the future detector development, to design the system so as to obtain the energy dependence of the neutron beam profile of the Back-n facility.

\section{The Micromegas detector}

\subsection{Detector design and setup}
\begin{figure*}[p]
\centering
  \includegraphics[width=\textwidth]{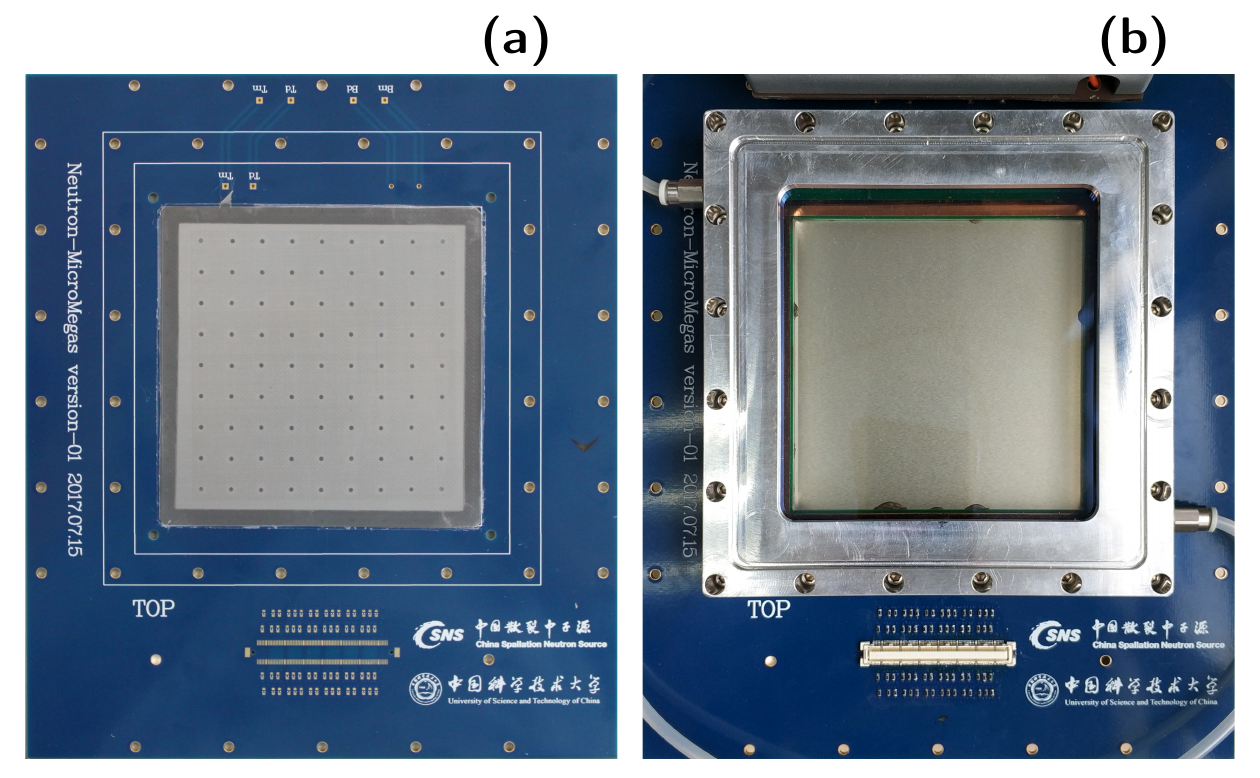}
  \caption{Micromegas fabrication process.  (a) The mesh after thermocompression bonding.  An array of cylindrical pillars is visible, which hold the mesh and ensure a constant distance between the mesh and the PCB.  (b) The detector chamber.}
  \label{fig:MMdet}
 \end{figure*}

\begin{figure*}[p]
\centering
  \includegraphics[width=\textwidth]{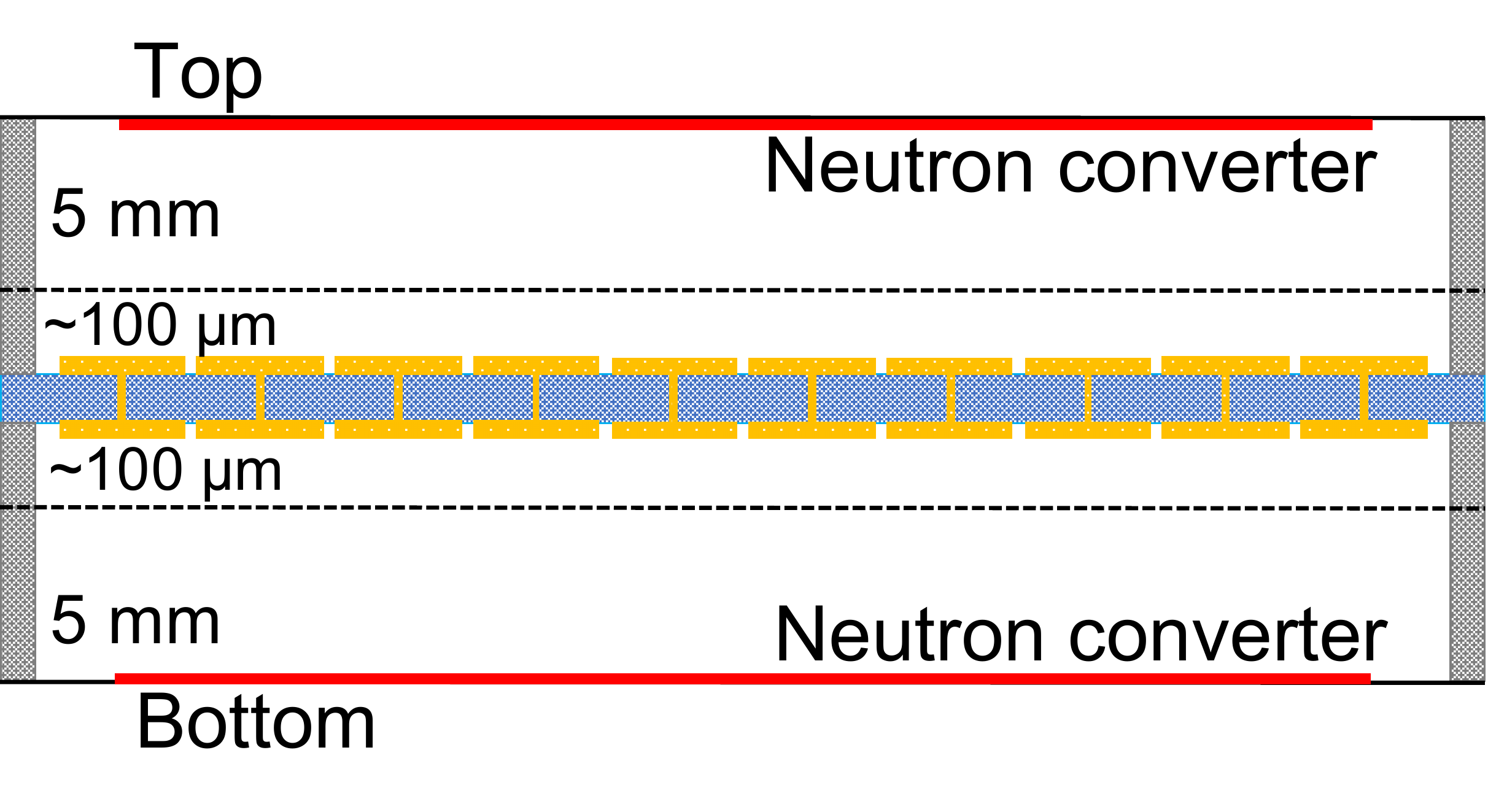}
  \caption{Schematic of the MMD with back-to-back double-avalanche structure.  The mesh (dashed line) separates the \SI{5}{\mm} drift gap from the \SI{100}{\um} avalanche gap.  Each of the XY readout strips is independently connected to the central board.}
  \label{fig:back2back}
\end{figure*}

Figure~\ref{fig:MMdet} shows the MMD of this work.  The neutron converter is attached to the cathode electrode facing the drift gap (i.e. the region between cathode and mesh).  The mesh-anode avalanche gap (an amplification region where electron avalanches occur) is manufactured with the thermal bonding method (see Ref.~\cite{bib:Feng2019} for detailed description of the manufacturing process).  With this method, each side of the readout PCB is attached to a metallic mesh to form the avalanche gap, and a back-to-back structure is obtained (see Fig.~\ref{fig:back2back}).  Hereafter, the top (bottom) part of the detector is defined as the upper (lower) half of the back-to-back structure shown in Fig.~\ref{fig:back2back}.  The mesh and the anode plane are separated by insulating pillars.  Each pillar has a cylindrical shape of \SI{2}{\mm} in diameter.  These two avalanche structures are able to work simultaneously for a higher detection efficiency.  The avalanche gap and the drift gap are \SI{100}{\um} and \SI{5}{\mm}, respectively.  The total active area of the MMD is \SI{90x90}{\mm}.

\begin{figure*}[p]
\centering
  \includegraphics[width=\textwidth]{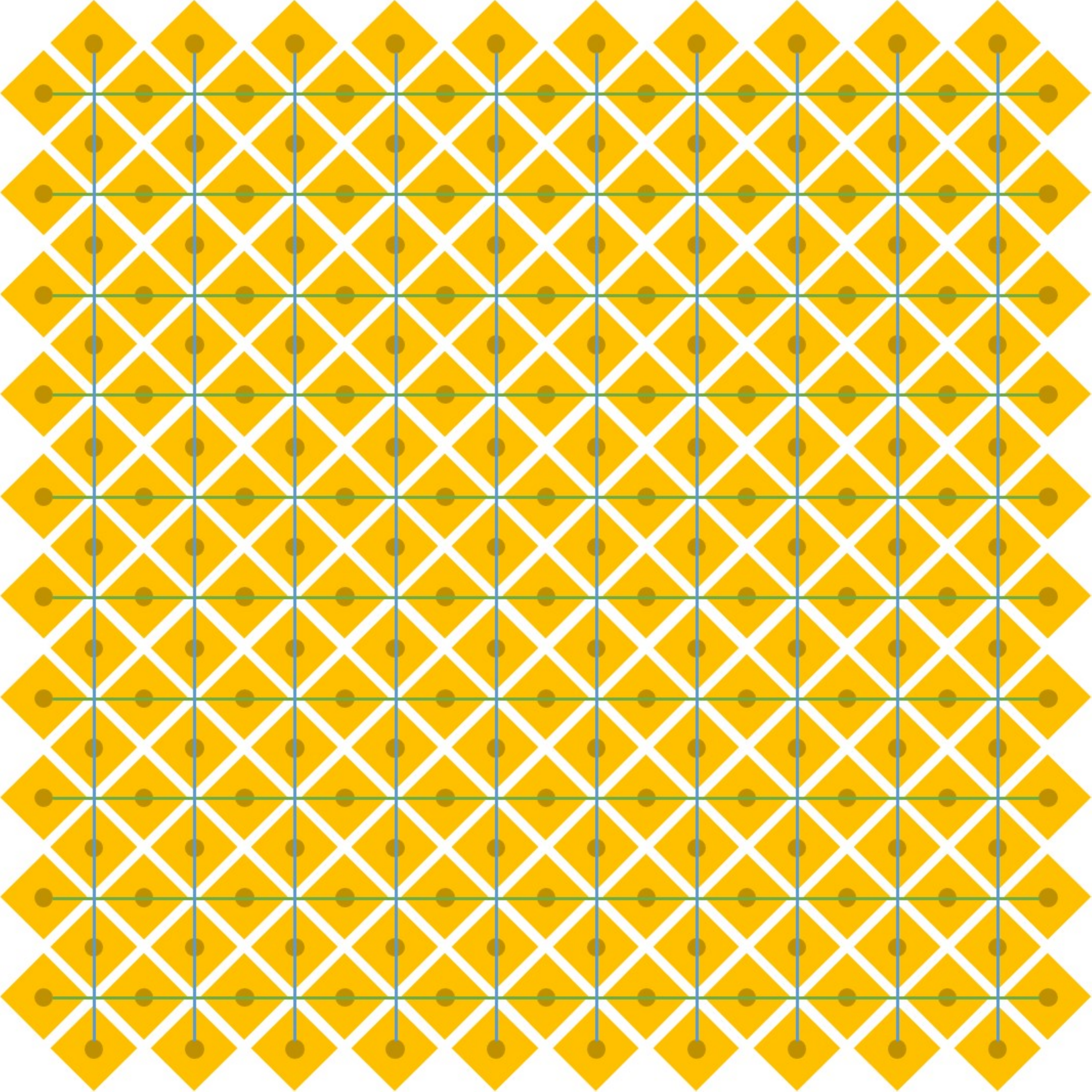}
  \caption{Schematics of the 2D readout structure.  The squares represent the copper pads that cover the surface of readout PCB.  The pads are interconnected by two groups of wires (represented by the horizontal and vertical lines) to the readout electronics through conductive vias in two inner layers of the readout PCB, such that orthogonal strip readouts are formed.}
  \label{fig:rd}
 \end{figure*}

Since the 2D position information of the particle hits is required for the neutron beam monitoring, a 2D readout structure has been designed and built for the MMD (see Fig.~\ref{fig:rd}) to reconstruct the 2D position using strip coincidences.  This readout structure allows a better determination of the two coordinates of the position from the charge produced in the amplification volume and induced on the anode strips.  A special readout PCB of 2 mm thick has been designed and adapted to the 2D readout strips shown in Fig.~\ref{fig:rd}.  Square copper pads, each with dimensions of \SI{0.96x0.96}{\mm}, are densely arranged on the surface of the readout PCB, with a bevel of \SI{45}{\degree}.  The pads are interconnected by two groups of wires in orthogonal directions, such that symmetrical XY strip readouts are formed, with 64 strips (\SI{1.5}{\mm} pitch) in each direction.  Each readout strip is capacitively coupled to a charge amplifier, corresponding to one readout channel.  The goals of this 2D readout scheme are (a) to mitigate the unequal charge sharing between the two strip layers that occurs in standard XY detectors and (b) to minimize the material budget of the detector.

Two detector units (namely MMD-MA and MMD-RA) with different anode designs are fabricated.  For MMD-MA, copper anode is used; for MMD-RA, resistive anode with germanium coating is used.  The copper-anode MMD-MA is easier to fabricate than the resistive-anode MMD-RA, but has lower gain than the latter.  The resistive layer of MMD-RA protects the detector from discharges caused by possible intense ionization.
 
\subsection{Front-end electronics}
A dedicated front-end electronics (FEE) system based on the AGET ASIC chip~\cite{bib:Li2017, bib:Li2018} is developed to process the MMD strip signals that are read out independently.  Each AGET chip features 64 analog channels.  In total, two AGET chips with 128 channels are used.  Each channel records 512 sampling points with adjustable sampling frequency from \SIrange{3}{100}{\MHz}, which meets the basic requirement for the signal data processing.  The strip signal is amplified by a charge sensitive preamplifier and a shaping amplifier, and subsequently digitized by a multi-channel analyzer. The raw signal of each readout channel is presented as a 12-bit ADC value.

\section{Performance tests with X-rays and $\alpha$ particles}

\subsection{Detector characterization with X-rays}
A ${}^{55}\mathrm{Fe}$ X-ray source ($E_{\mathrm{K_{\alpha}}} = \SI{5.9}{\keV}$, $E_{\mathrm{K_{\beta}}} = \SI{6.5}{\keV}$) was used to test the detector performance, including the relative electron transparency, the gain, and the gain uniformity.  The ${}^{55}\mathrm{Fe}$ source was collimated by a \SI{3}{\mm} diameter pinhole.  The detector chamber was filled with a gas mixture of 93\% argon (Ar) and 7\% $\mathrm{CO}_2$ at atmospheric pressure.

\begin{figure*}[p]
\centering
  \includegraphics[width=\textwidth]{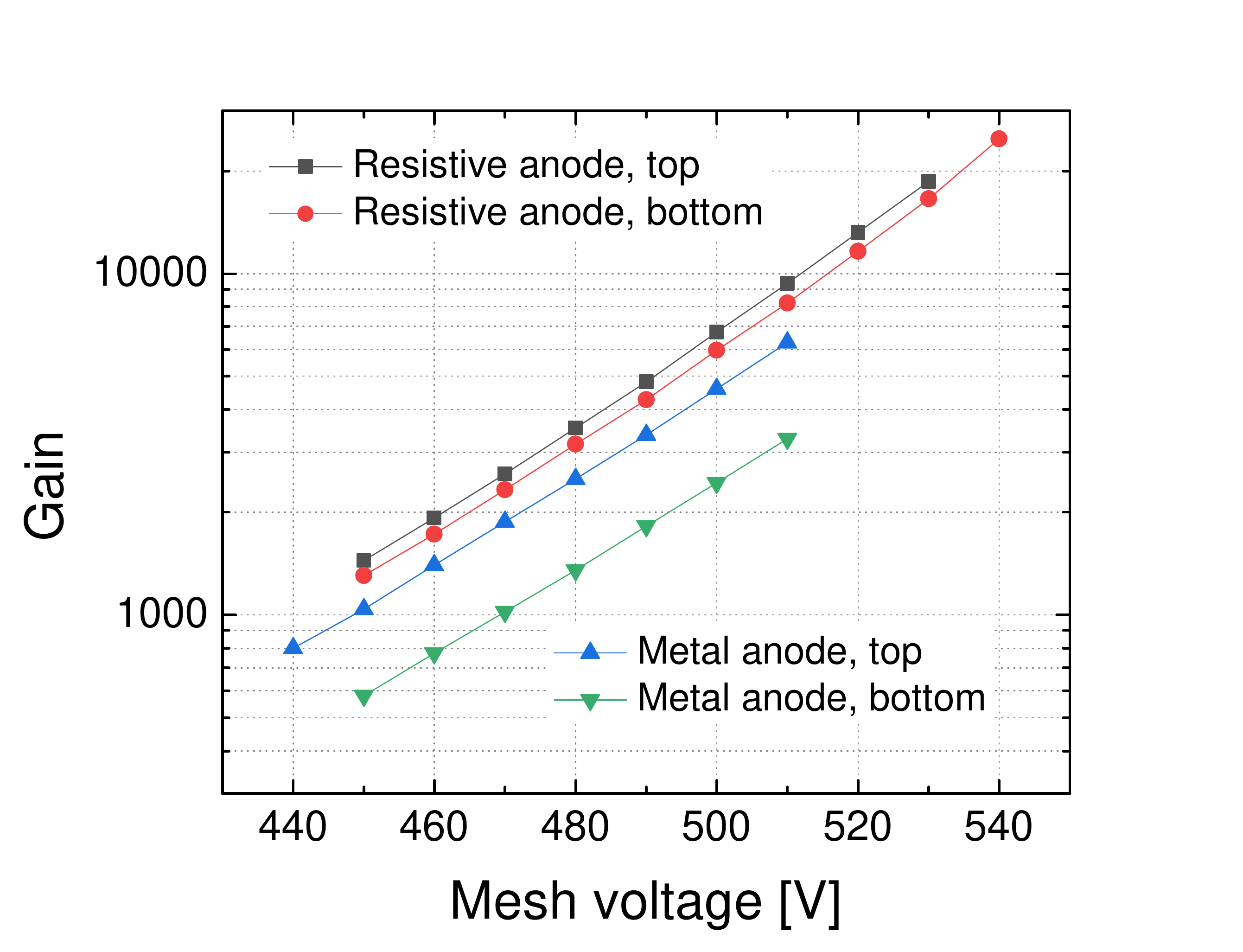}
  \caption{Measured gains of MMD-RA (squares and circles) and MMD-MA (upward- and downward-pointing triangles) as a function of the voltage applied to the mesh with the $E_\mathrm{mesh} / E_\mathrm{drift}$ ratio fixed to 160.}
  \label{fig:gain}
 \end{figure*}

The relative electron transparency for both MMD-MA and MMD-RA was found to reach a plateau when the ratio of the electric fields in the avalanche gap and in the drift gap ($E_\mathrm{mesh} / E_\mathrm{drift}$) was greater than 150.  Therefore in the subsequent performance studies, the $E_\mathrm{mesh} / E_\mathrm{drift}$ ratio was fixed to 160 for a maximal electron transparency.  The measured gains of MMD-MA and MMD-RA are shown in Fig.~\ref{fig:gain}.  One can see that the highest gains of MMD-MA and MMD-RA (with $E_\mathrm{mesh}$ below the breakdown voltage) are about $20\,000$ and $3000$, respectively, which meet the basic requirement for the detection of the charged particles from the neutron interaction.  

\begin{figure*}[p]
\centering
  \includegraphics[width=\textwidth]{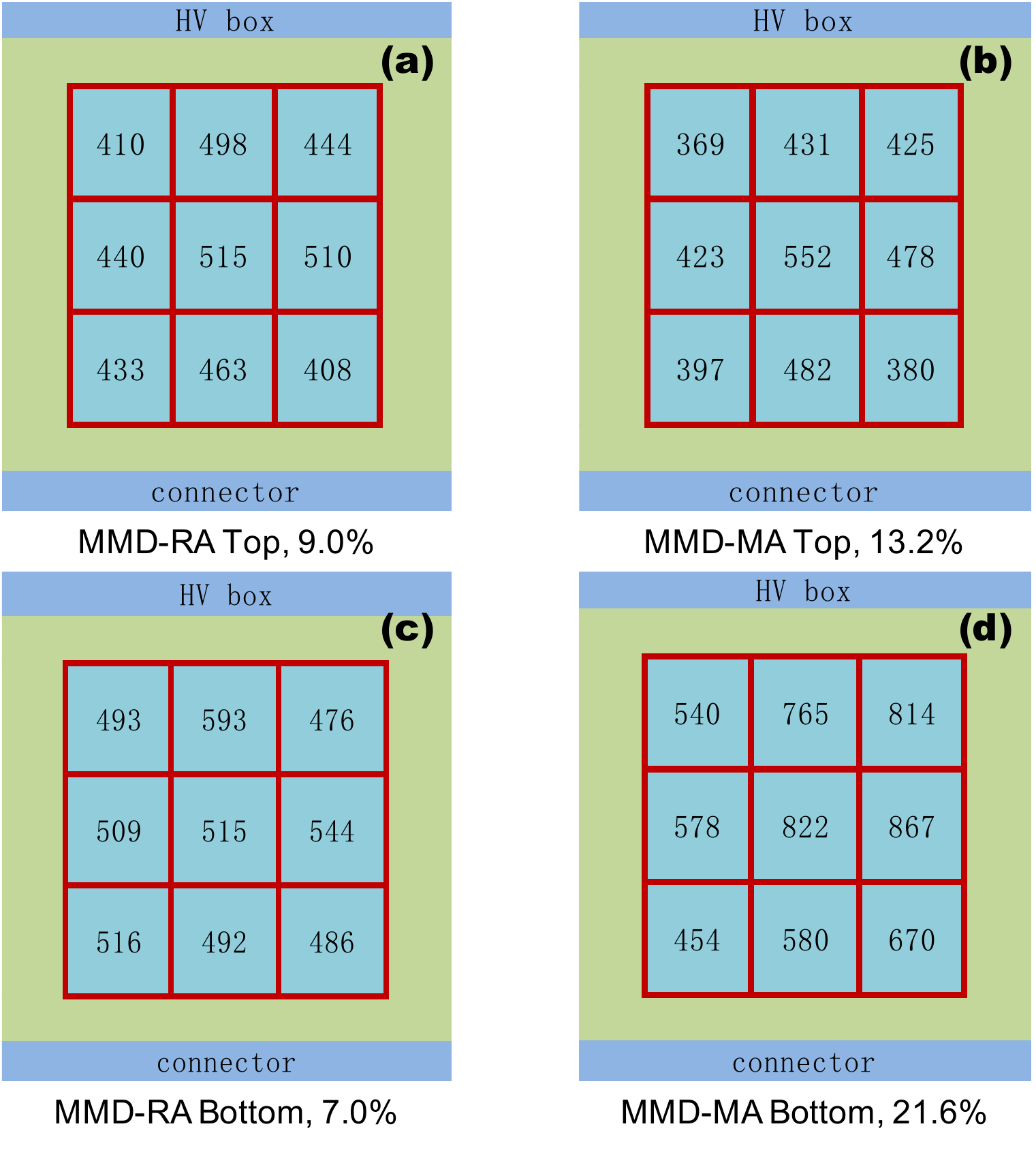}
  \caption{The number in the \SI{30x30}{\mm} cell indicates the position of the ${}^{55}\mathrm{Fe}$ dominant peak (in units of ADC channels) measured at the center of each cell for (a) the top part of MMD-RA, (b) the top part of MMD-MA, (c) the bottom part of MMD-RA, and (d) the bottom part of MMD-MA. The corresponding gain uniformity is calculated to be 9.0\%, 13.2\%, 7.0\%, and 21.6\%, respectively.}
  \label{fig:gain_uni}
 \end{figure*}

The absolute gain was extracted from the ratio of the charge collected from the anode corresponding to the ${}^{55}\mathrm{Fe}$ \SI{5.9}{\keV} full energy peak to the charge produced by the primary ionization.  This charge (collected on the anode strips) is proportional to the recorded signal amplitude.  The gain uniformity was calculated as the standard deviation of the gain values measured at different positions of the detector active area normalized to the mean.  In total, the gains were measured at nine areas (each of \SI{30x30}{\mm}) using the ${}^{55}\mathrm{Fe}$ source collimated to \SI{3}{\mm} in diameter to irradiate the central part of each area (see Fig.~\ref{fig:gain_uni}).  Overall, the gain uniformities of the top and bottom detector part of MMD-MA were measured to be 13\% and 22\%, respectively.  With the optimization of the thermal bonding method, the top and bottom uniformities of MMD-RA were improved to 9\% and 7\%, respectively.  As can be seen in Fig.~\ref{fig:gain_uni}, the gains at the edges of the detector are systematically lower.  This will be optimized by further technical improvements of the thermal bonding method.

Using the collimated ${}^{55}\mathrm{Fe}$ source, the energy resolution for MMD-MA and MMD-RA was measured to be 37\% and 28\% (FWHM) at \SI{5.9}{\keV}, respectively.  This is worse than the energy resolution of 16\% presented in Ref.~\cite{bib:Feng2019} mainly because of the the anode structure.  Since the gap between the anode pads is \SI{0.1}{\mm}---comparable to the avalanche gap length---the effects of the electric field inhomogeneities near the edges of the pads are nonnegligible.  This leads to considerable gain variations and consequently deteriorates the energy resolution.

\subsection{Position reconstruction test with an $\alpha$ source}
\begin{figure*}[p]
\centering
  \includegraphics[width=\textwidth]{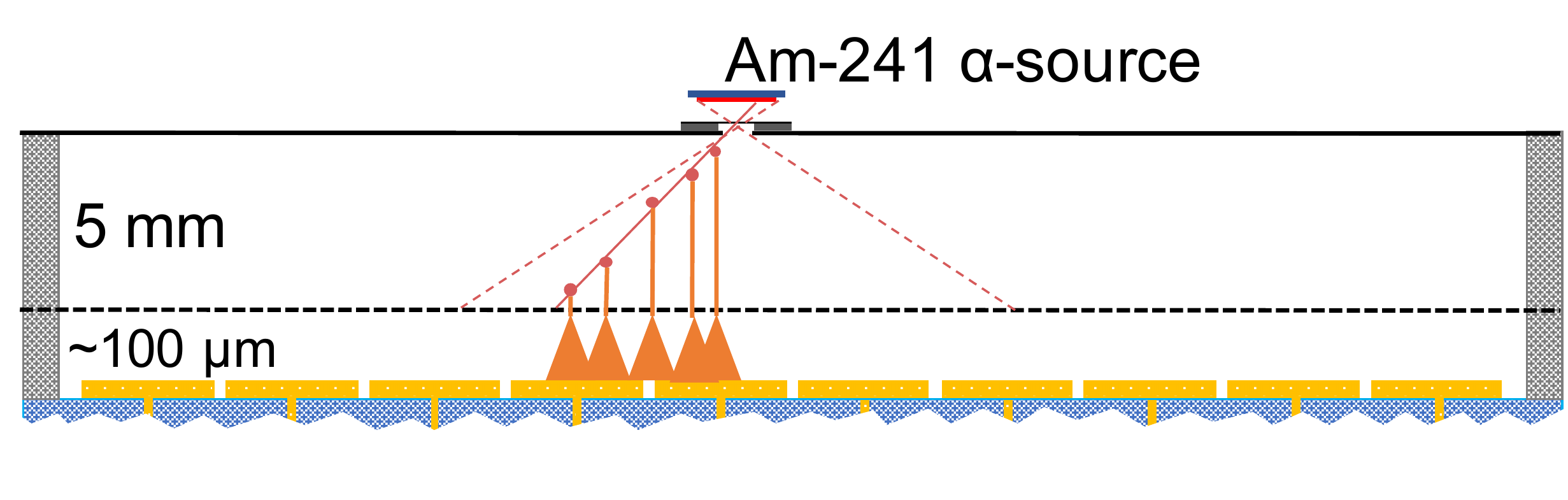}
  \caption{Schematic of the setup for the position reconstruction test with the ${}^{241}\mathrm{Am}$ $\alpha$ source.}
  \label{fig:Am_test}
 \end{figure*}

The reconstruction of the neutron interaction position is a crucial part of the neutron beam profiling with MMD.  Position reconstruction algorithms were tested with an ${}^{241}\mathrm{Am}$ source which emits $\alpha$ particles in order to choose the best one for the reconstruction of the neutron interaction point at the measurements with neutron beams.  As shown in Fig.~\ref{fig:Am_test}, the \SI{\sim5.4}{\MeV} $\alpha$ particles from the ${}^{241}\mathrm{Am}$ source were collimated by a \SI{1}{\mm} diameter pinhole on the drift cathode, and subsequently attenuated by the air and a \SI{\sim10}{\um} thick plastic film before entering the drift gap.  MMD-MA was used for this test and only one side of the detector was operating during this test.
 
\begin{figure*}[p]
\centering
  \includegraphics[width=\textwidth]{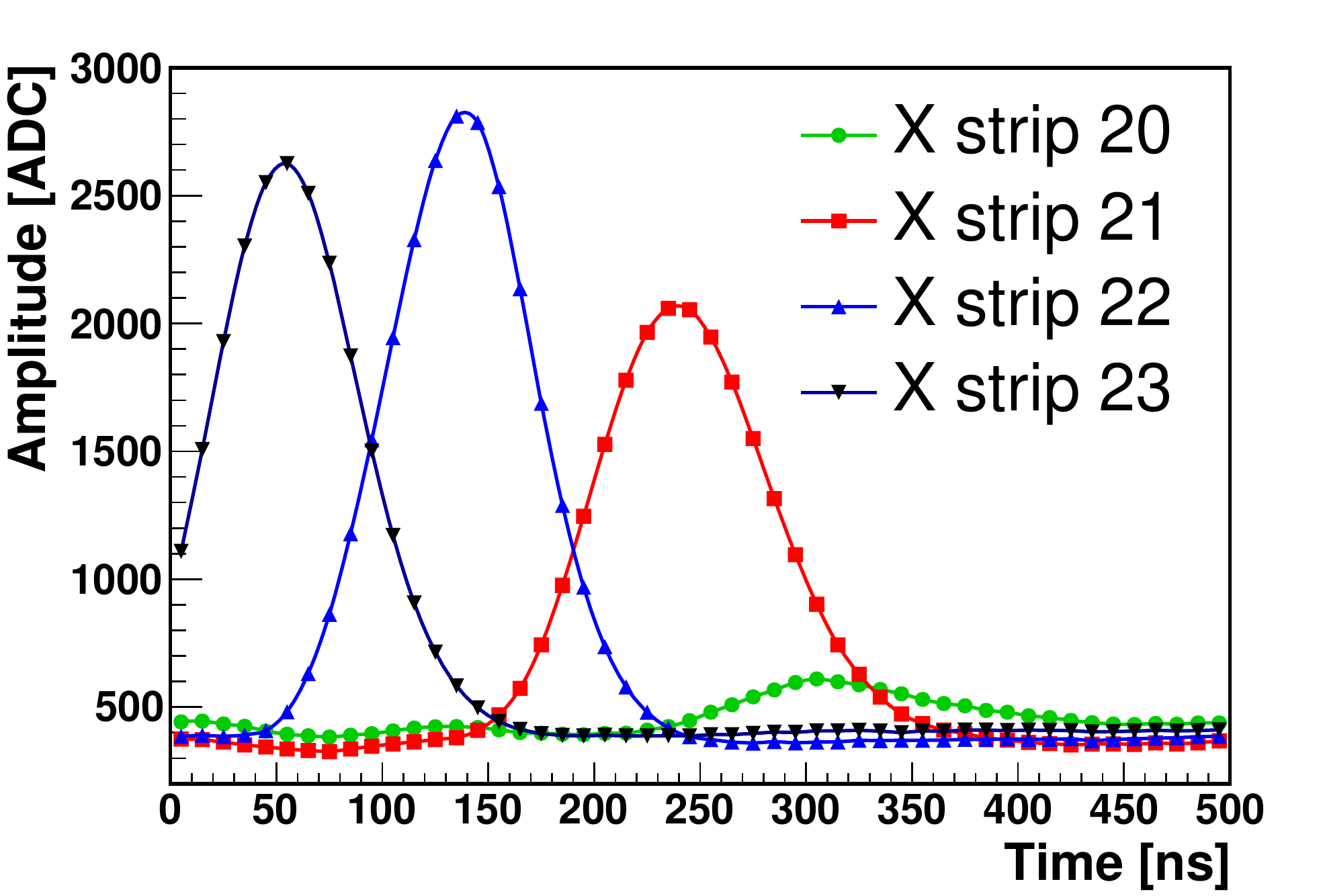}
  \caption{Raw signal waveforms from four consecutive readout strips recorded from an $\alpha$ event.  The sampling frequency is \SI{100}{\MHz}.}
  \label{fig:waveform_alpha}
 \end{figure*}

Two different algorithms for the hit position reconstruction were studied: (a) the charge centroid method, which reconstructs the hit position as the average of the strip position weighted by the strip charge, and (b) the micro Time Projection Chamber ($\mu$TPC) method~\cite{bib:Badertscher2011,bib:Kubo2003,bib:Nishimura2006}, which exploits the ability of the MMD to operate as a $\mu$TPC.  In the latter method, the (relative) arrival time of ionization charge on the strip is measured, and the last strip that has a signal is considered as an estimate of the neutron interaction position in the converter.  In principle, the centroid method is accurate for the reconstruction of interaction position of perpendicular neutron beams, but the resolution deteriorates with increasing angle of the secondary particles produced from the converter, since the signal extends over a larger number of strips.  By using the $\mu$TPC method, the origin of the primary ionization in the drift gap can be determined, which is a more precise estimate of the neutron interaction position than the charge centroid.  Since the peaking time of the FEE can be as short as \SI{70}{\ns}, it allows a simultaneous and independent recording of the time information of the strips.  Figure~\ref{fig:waveform_alpha} shows the waveforms of the raw signals from four consecutive strips.  As can be seen, the time differences between adjacent strips are evident.

\begin{figure*}[p]
\centering
  \includegraphics[width=\textwidth]{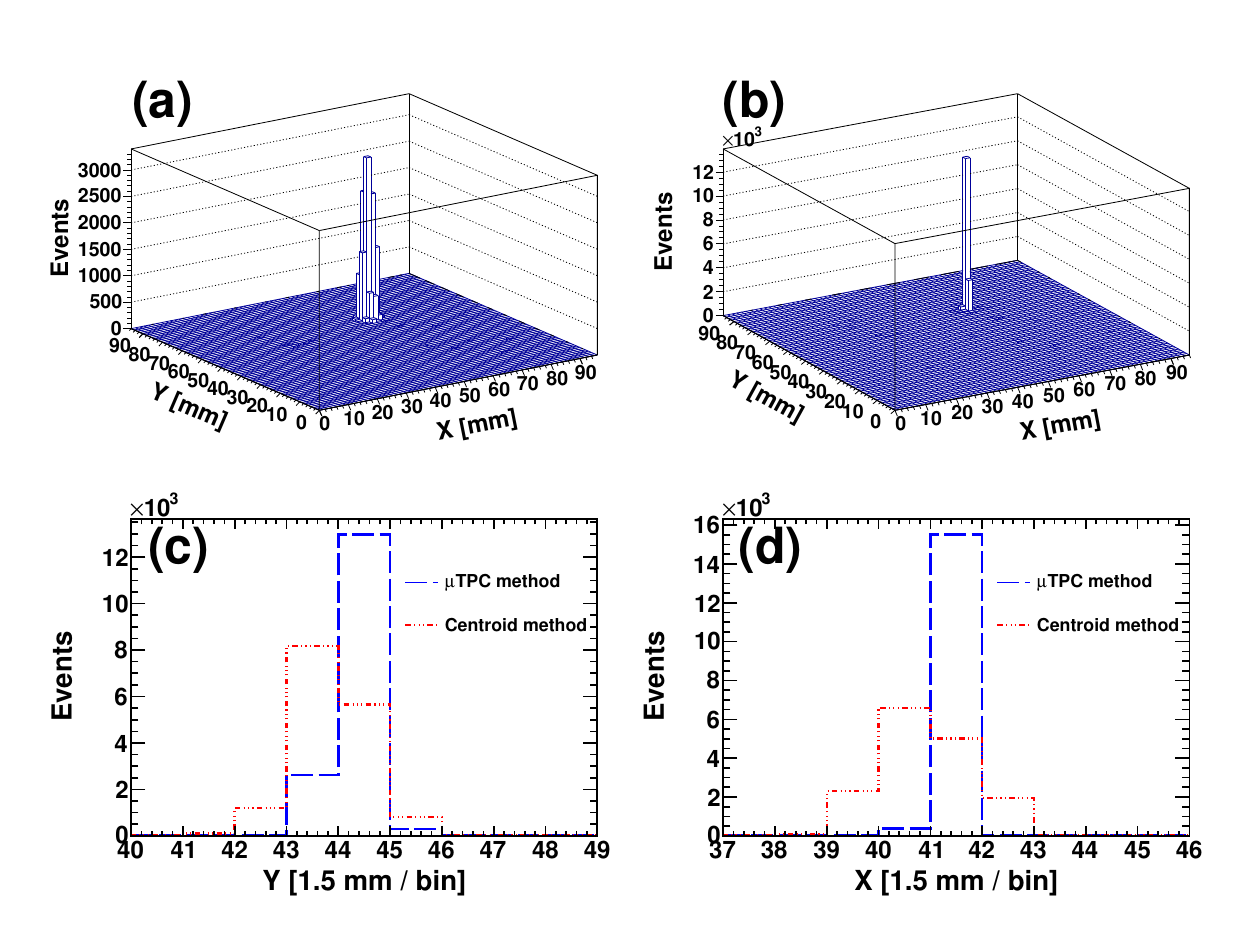}
  \caption{Distributions of the reconstructed position using (a) the charge centroid method and (b) the $\mu$TPC method, and their 1D projections to (c) the Y-axis and (d) the X-axis.}
  \label{fig:hitpos}
 \end{figure*}

In order to compare the performance of these two algorithms, the $\alpha$ events are required to have at least two fired strips.  Figure~\ref{fig:hitpos} shows the two coordinates of the position where the $\alpha$ particles exiting the pinhole collimator reconstructed by these two methods, along with the 1D projections.  The distribution of the reconstructed position obtained from the $\mu$TPC method has a much narrower width than that from the charge centroid method, indicating a better spatial resolution for the $\mu$TPC method.

\section{Neutron beam spot distribution measurement}

\subsection{Neutron converter}
\begin{figure*}[p]
\centering
  \includegraphics[width=\textwidth]{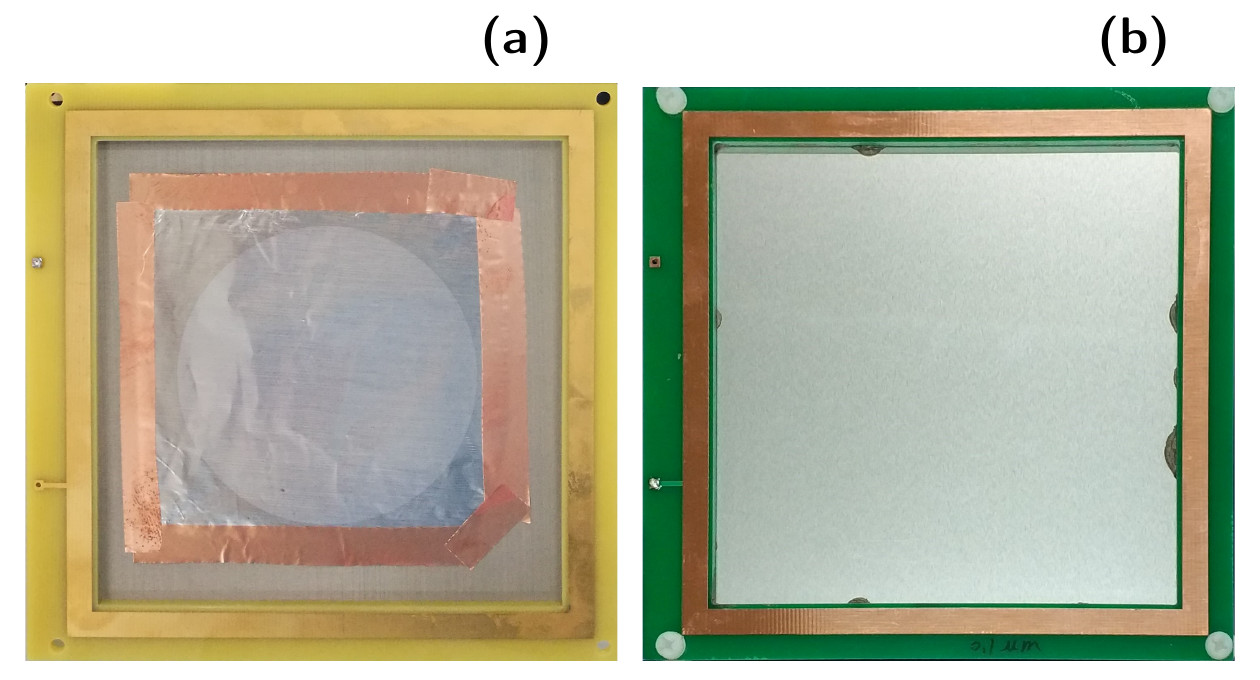}
  \caption{(a) The neutron converter consisting of a ${}^{6}\mathrm{LiF}$ layer of \SI{6}{\cm} diameter and \SI{3}{\um} thickness deposited on the surface of the \SI{\sim10}{\um} thick aluminum foil facing the drift gap.  A metallic mesh is glued to the other side of the aluminum foil by conductive copper foil tapes to form the drift electrode.  (b) The neutron converter consisting of a \SI{9x9}{\cm}\,$\times$\,\SI{0.1}{\um} natural boron layer deposited on the entire surface of the \SI{\sim10}{\um} thick aluminum foil drift electrode facing the drift gap.}
  \label{fig:nconv}
 \end{figure*}

\begin{figure*}[p]
\centering
  \includegraphics[width=\textwidth]{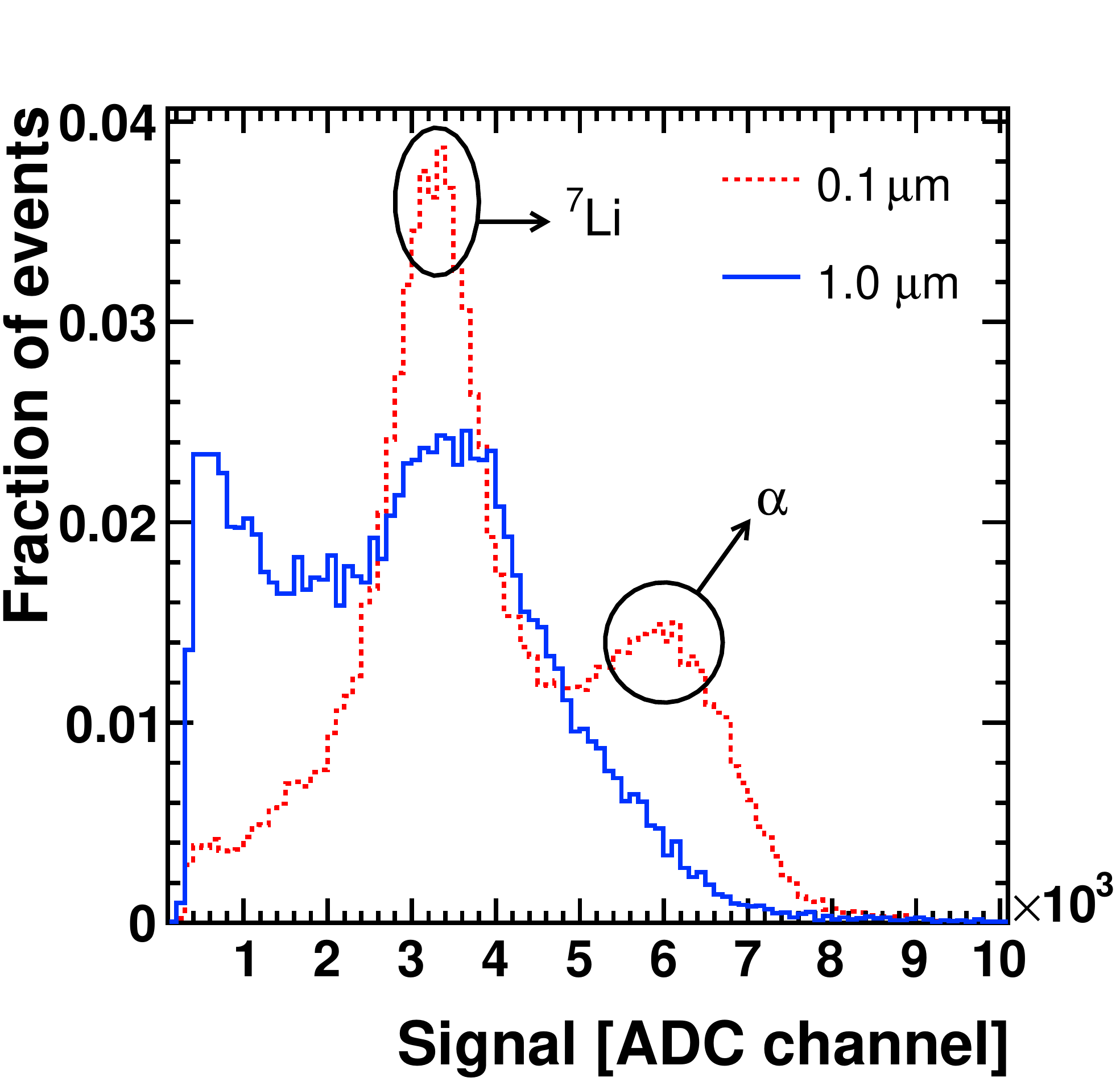}
  \caption{Distribution of the total amplitude obtained by summing up the amplitudes of all the anode signals for MMD-RA-B0.1 (red dashed line) and MMD-RA-B1.0 (blue solid line). The separation of the two reaction products, ${}^{7}\mathrm{Li}$ and $\alpha$, is better for the \SI{0.1}{\um} thick ${}^{10}\mathrm{B}$ converter.}
  \label{fig:10Bcomp}
 \end{figure*}

The neutrons are indirectly detected by the detection of the secondary charged particles produced in the interaction of neutrons with a target (neutron converter) deposited on the drift electrode.  Two types of neutron converters with well-known cross sections (${}^{6}\mathrm{Li}$ and ${}^{10}\mathrm{B}$) were used.  When ${}^{6}\mathrm{Li}$ is used, tritons and $\alpha$ particles are produced via the ${}^{6}\mathrm{Li} (n,\,t) {}^{4}\mathrm{He}$ reaction~\cite{bib:McGregor2003}:
\begin{eqnarray}
n+{}^{6}\mathrm{Li} \rightarrow \alpha\,(\SI{2.05}{\MeV})+t\,(\SI{2.73}{\MeV}).
\label{eq:nLi6}
\end{eqnarray}
Figure~\ref{fig:nconv}~(a) shows a neutron converter which consists of a ${}^{6}\mathrm{LiF}$ layer of \SI{6}{\cm} diameter and \SI{3}{\um} thickness deposited on the surface of the \SI{\sim10}{\um} thick aluminum foil drift electrode facing the drift gap.  When ${}^{10}\mathrm{B}$ is used, it undergoes the ${}^{10}\mathrm{B} (n,\,\alpha) {}^{7}\mathrm{Li}$ reaction:
\begin{equation}
n+{}^{10}\mathrm{B}\rightarrow{}^{7}\mathrm{Li}\,(\SI{0.84}{\MeV}) + \alpha\,(\SI{1.47}{\MeV}) + \gamma\,(\SI{0.48}{\MeV}),
\label{eq:nB10_1}
\end{equation}
or
\begin{equation}
n+{}^{10}\mathrm{B}\rightarrow{}^{7}\mathrm{Li}\,(\SI{1.01}{\MeV}) + \alpha\,(\SI{1.78}{\MeV}),
\label{eq:nB10_2}
\end{equation}
for which the branching fractions are 94\% and 6\%, respectively.  The energies of the reaction products are shown in Eqs.~\eqref{eq:nLi6},~\eqref{eq:nB10_1},~and~\eqref{eq:nB10_2} for the cases of thermal neutrons.  Figure~\ref{fig:nconv}~(b) shows a neutron converter which consists of a natural boron layer (19.9\% ${}^{10}\mathrm{B}$) of \SI{9x9}{\cm}\,$\times$\,\SI{0.1}{\um} deposited over the entire surface of the \SI{\sim10}{\um} thick aluminum foil drift electrode facing the drift gap.  In addition, a thicker ${}^{10}\mathrm{B}$ layer (\SI{1.0}{\um}) was also produced and used with the MMD for performance comparison with the \SI{0.1}{\um} one.  Figure~\ref{fig:10Bcomp} shows the performance comparison between these two ${}^{10}\mathrm{B}$ converters used with MMD-RA (namely MMD-RA-B0.1 and MMD-RA-B1.0), based on the data collected in a test at Back-n during the beam commissioning.  As can be seen, the two reaction products ${}^{7}\mathrm{Li}$ and $\alpha$ are well separated for the thinner ${}^{10}\mathrm{B}$ converter.

\subsection{Image reconstruction with an Am-Be neutron source}
\begin{figure*}[p]
\centering
  \includegraphics[width=\textwidth]{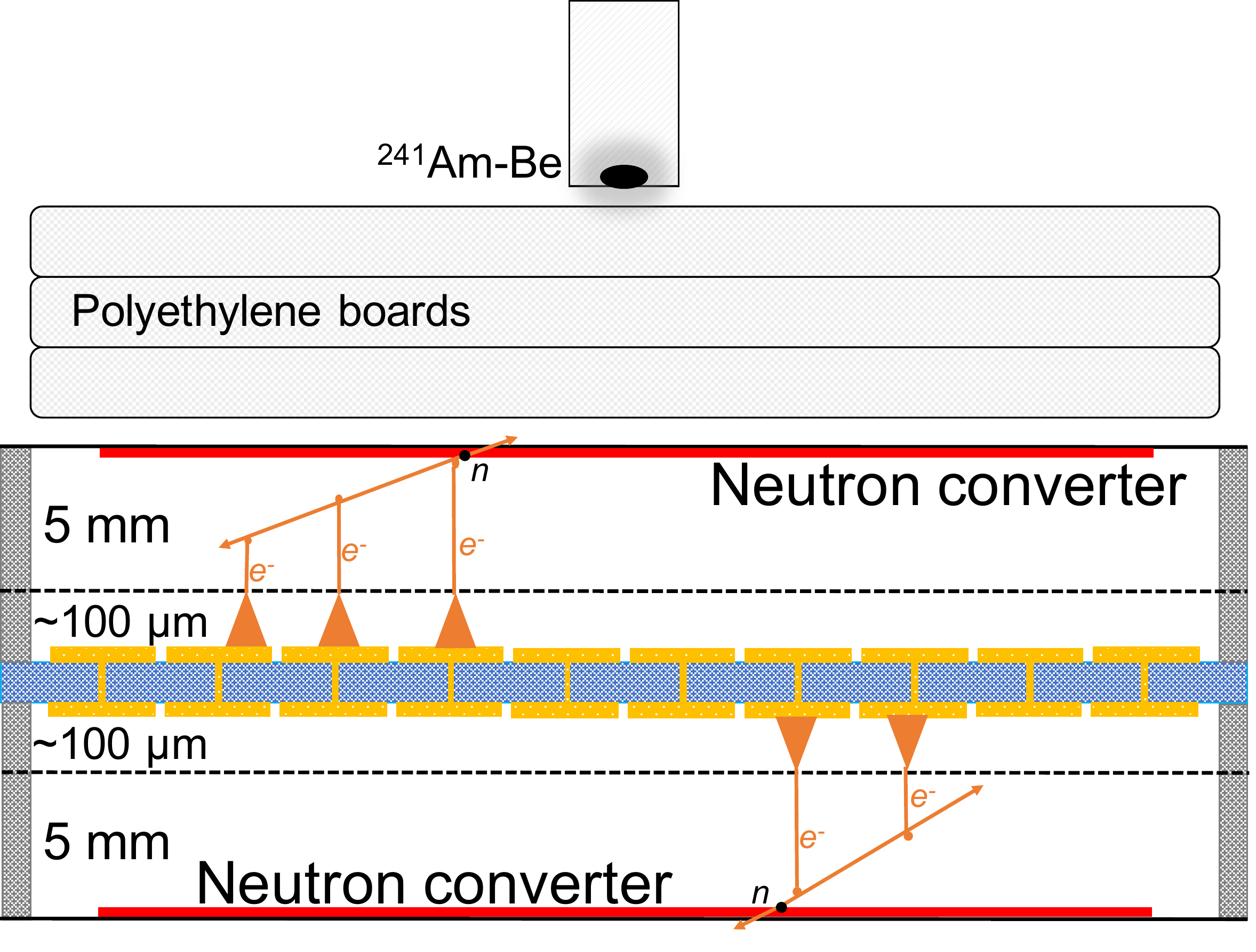}
  \caption{Schematic of the setup for the detector performance test with the Am-Be neutron source.}
  \label{fig:Am-Be_test}
 \end{figure*}

The performance of the MMD at the image reconstruction was tested with an Am-Be neutron source prior to the measurement of the Back-n neutron beam profile.  As shown in Fig.~\ref{fig:Am-Be_test}, MMD-MA with ${}^{6}\mathrm{Li}$ converter (MMD-MA-Li) was irradiated with the neutrons from the Am-Be neutron source.  In order to moderate these neutrons (with \SI{\sim4.5}{\MeV} mean energy), three \SI{1}{\cm} thick polyethylene layers were placed between the source and MMD-MA-Li. 

\begin{figure*}[p]
\centering
  \includegraphics[width=\textwidth]{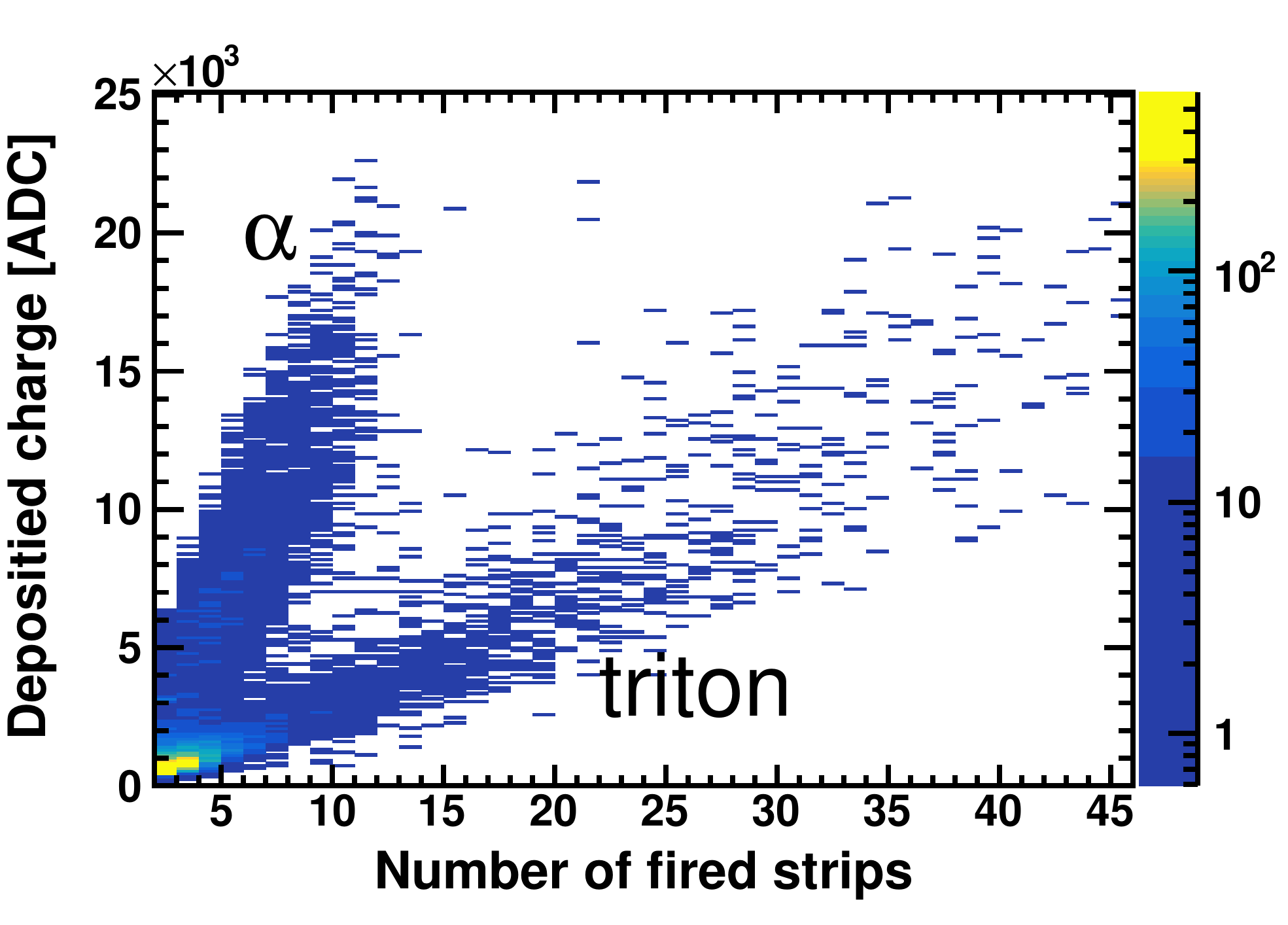}
  \caption{Total charge deposited at the strips as a function of the total number of those fired strips.}
  \label{fig:PID}
 \end{figure*}

Figure~\ref{fig:PID} shows the total charge deposited at the X and Y strips as a function of the total number of those fired strips.  Two bands can be seen, corresponding to the two reaction products, the tritons and the $\alpha$ particles, from the ${}^{6}\mathrm{Li} (n,\,t) {}^{4}\mathrm{He}$ reaction.  According to the Bethe-Bloch formula, the ionization energy loss per unit path length is proportional to $Z^2$, where $Z$ is the charge of the incident particle.  Therefore the $\alpha$ particles have a shorter range than the tritons in the drift gap, which results in low strip multiplicities for the $\alpha$ tracks, i.e. the ionization energy deposition of the $\alpha$ tracks was only recorded by few strips near the neutron interaction point.  Thus, in Fig.~\ref{fig:PID} the band on the left corresponds to the $\alpha$ events, while the band on the right corresponds to the triton events.

\begin{figure*}[p]
\centering
  \includegraphics[width=\textwidth]{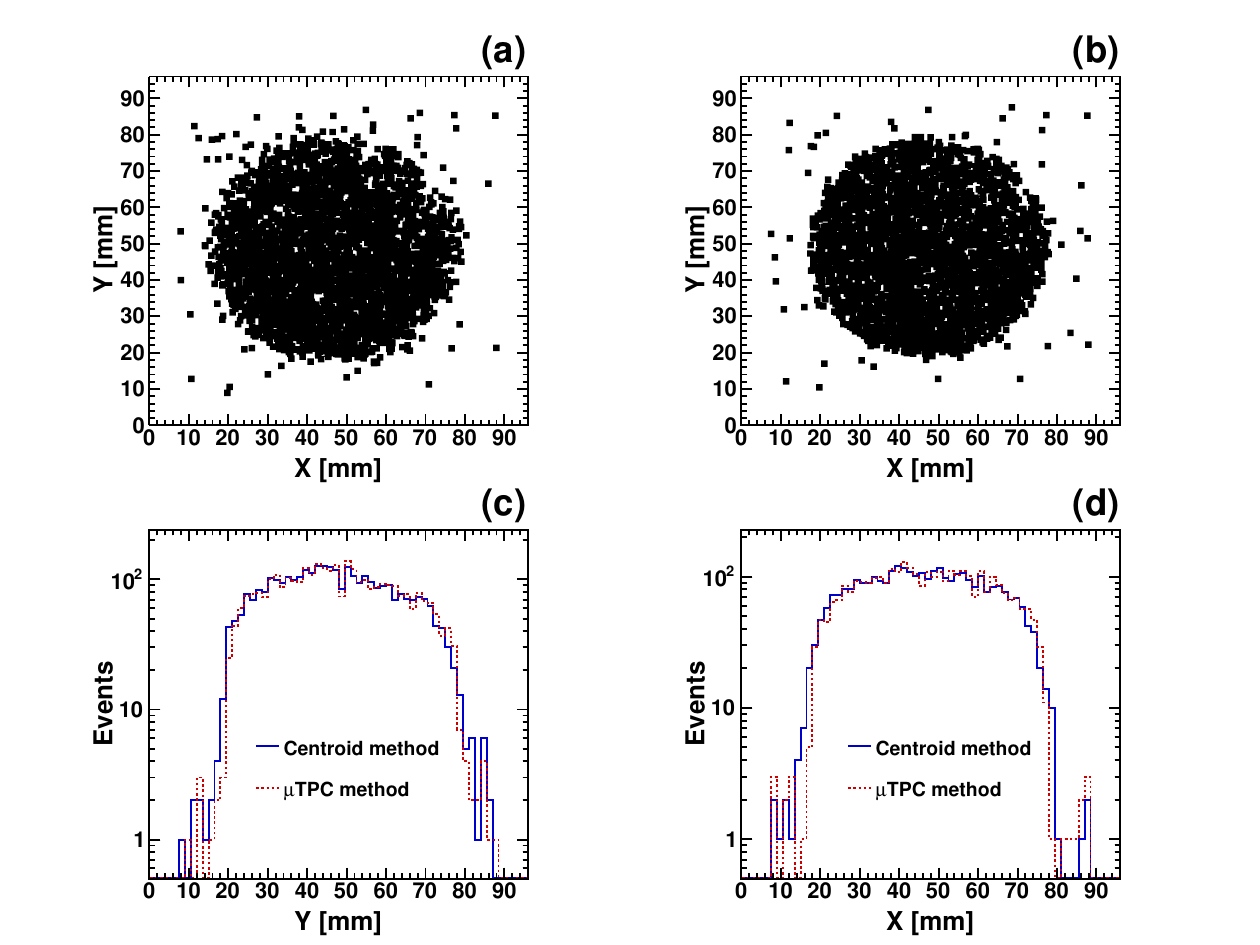}
  \caption{Positions of interaction of the neutrons with the ${}^{6}\mathrm{LiF}$ layer reconstructed by (a) the charge centroid method and (b) the $\mu$TPC method, and their 1D projections to (c) the Y-axis and (d) the X-axis.}
  \label{fig:Am-Be_prof}
 \end{figure*}

The image of the ${}^{6}\mathrm{LiF}$ neutron converter was reconstructed using the charge centroid method and the $\mu$TPC method, as shown in Fig.~\ref{fig:Am-Be_prof}~(a) and Fig.~\ref{fig:Am-Be_prof}~(b), respectively.  Again, a clearer edge of the 2D reconstructed image was obtained by using the $\mu$TPC method.  This observation can be noticed in Fig.~\ref{fig:Am-Be_prof}~(c) and Fig.~\ref{fig:Am-Be_prof}~(d), where sharper edges of the 1D projections were obtained with the $\mu$TPC method.

\subsection{Neutron beam spot distribution measurement at Back-n}
\begin{figure*}[p]
\centering
  \includegraphics[width=\textwidth]{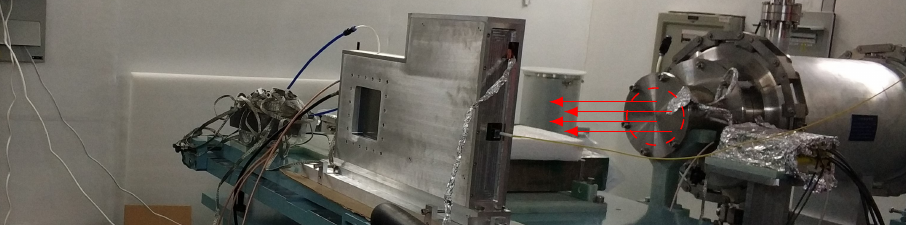} 
  \caption{The MMD together with the FEE system was placed in the shielding container made of aluminum, as a neutron beam profiler at Back-n.  The arrows represent the incoming neutron beam.}
  \label{fig:exp_setup}
 \end{figure*}

MMD-MA with \SI{1.0}{\um} thick ${}^{10}\mathrm{B}$ converter (MMD-MA-B1.0) was placed at Back-n to study the performance of neutron detection as well as the reconstruction of neutron beam profiles.  Along the Back-n white neutron beam line there are two experimental halls, Endstation-1 (ES\#1) and Endstation-2 (ES\#2), which are about \SI{55}{\m} and \SI{76}{\m} away from the spallation target, respectively (Fig. 4 of Ref.~\cite{bib:Zhang2018}).  The detailed description of Back-n is presented in Ref.~\cite{bib:Zhang2018}.  The detector was placed at a location in ES\#1 which is about \SI{56}{\m} away from the spallation target, and a picture of the experimental setup is shown in Fig.~\ref{fig:exp_setup}.   During the data taking period, the CSNS accelerator was operating stably at a repetition rate of \SI{25}{\Hz} and a proton beam power of \SI{20}{\kW}.

Only the top part of the MMD-MA-B1.0 was operating during this measurement.  The thicker ${}^{10}\mathrm{B}$ converter was chosen in this case for a higher neutron reaction rate.  The detection gas used was 90\% Ar and 10\% $\mathrm{CO}_2$ at atmospheric pressure.  Limited by the gas flow system at ES\#1, a slightly lower fraction of Ar was used compared to the previous tests.  This setting is within the working range of the detector and the resulting difference in the detector performance is negligible.  The dynamic range of the FEE and the sampling frequency of the signal recording were set to \SI{240}{\femto\coulomb} and \SI{5}{\MHz}, respectively.

The beam spot size at ES\#1 is determined by the apertures of the shutter and Collimator-1 of Back-n (see Fig.~6 and Table~1 of Ref.~\cite{bib:Zhang2018}).  For this measurement, the configuration of the collimation aperture was set to $\diameter\SI{50}{\mm}$ for the shutter and $\diameter\SI{50}{\mm}$ for Collimator-1, and the corresponding physics data collection time was about nine hours.  Under this configuration, the beam spot size at ES\#1 is expected to be about $\diameter\SI{50}{\mm}$ according to the Monte Carlo (MC) simulation study~\cite{bib:Zhang2018}.  A dedicated run without beam in the Back-n beam pipe (i.e. the shutter being closed) was taken for \SI{\sim230}{\s} to collect data for pedestal correction (see Section~\ref{sec:ped}).

\subsection{Data analysis}
\subsubsection{Pedestal calculation} \label{sec:ped}
Pedestal levels and noise values are calculated for each strip, using the physics run data as well as the pedestal run data for an independent cross-check.  For a raw signal waveform recorded by the readout channel $i$ for the event $j$, the ADC value of each sampling point, $R_{i,j}$, consists of four components: $R_{i,j}=P_i + CM_j + N_{i,j} + S_{i,j}$, where $i$ is the strip or channel number, $j$ is the event number, and $R_{i,j}$ is an integer between 0 and 4095.  The first component is the average pedestal level of strip $i$ for $n$ recorded events, which is given by $P_i=(1/n)\sum^n_{j=1}R_{i,j}$.  The second component is the so-called common mode of the AGET chip to which strip~$i$ belongs, calculated per-event as the mean of $R_{i,j}-P_i$ over 64 channels of the AGET chip.  The common mode is the deviation of all the channels of the AGET chip at the same time, mainly originating from the fluctuation of the reference ground of ADCs and differs on an event-by-event basis.  The last two components are the intrinsic noise and the signal, respectively.  The average intrinsic noise $\tilde{N}_i$ for strip $i$ is calculated as the standard deviation of the common-mode subtracted data.

When using data from a pedestal run to calculate the pedestal, the signal component is negligible.  However, when using data from a physics run to calculate the pedestal, the contribution from the signal component (in addition to the pedestal levels) in the raw physics data is nonnegligible.  To reduce the contamination from the signal, those $R_{i,j}$ values in the raw waveform satisfying $R_{i,j}-\tilde{P}_{i,j} \ge 3 \tilde{N}_{i,j}^{\prime}$ are excluded in the pedestal calculation, where $\tilde{P}_{i, j}$ and $\tilde{N}_{i,j}^{\prime}$ are the pedestal and the raw noise (i.e. calculated without the common-mode subtraction), respectively, both calculated up to event $j$.  It is worth to be noted that the results of the pedestal, common-mode, and noise calculation obtained from the physics run data are consistent with those obtained from the pedestal run data.

\subsubsection{Event selection}
The first step in the physics data analysis is the pedestal and common-mode subtraction.  After that, event selection criteria are applied for the selection of good neutron-induced charged particle events (hereafter referred to as neutron events) and for the rejection of background events or not well recorded neutron events.  First, strips with $A_i/\tilde{N}_i < 5$ are rejected for the event, where $A_i$ is the amplitude of the digitized waveform for strip $i$ after the pedestal and common-mode subtraction.  Second, if two or more strips are fired and recorded in an event, those strips should be consecutive, which ensures a continuous ionization of the gas when the charged particle traverses the drift gap.

\begin{figure*}[p]
\centering
  \includegraphics[width=\textwidth]{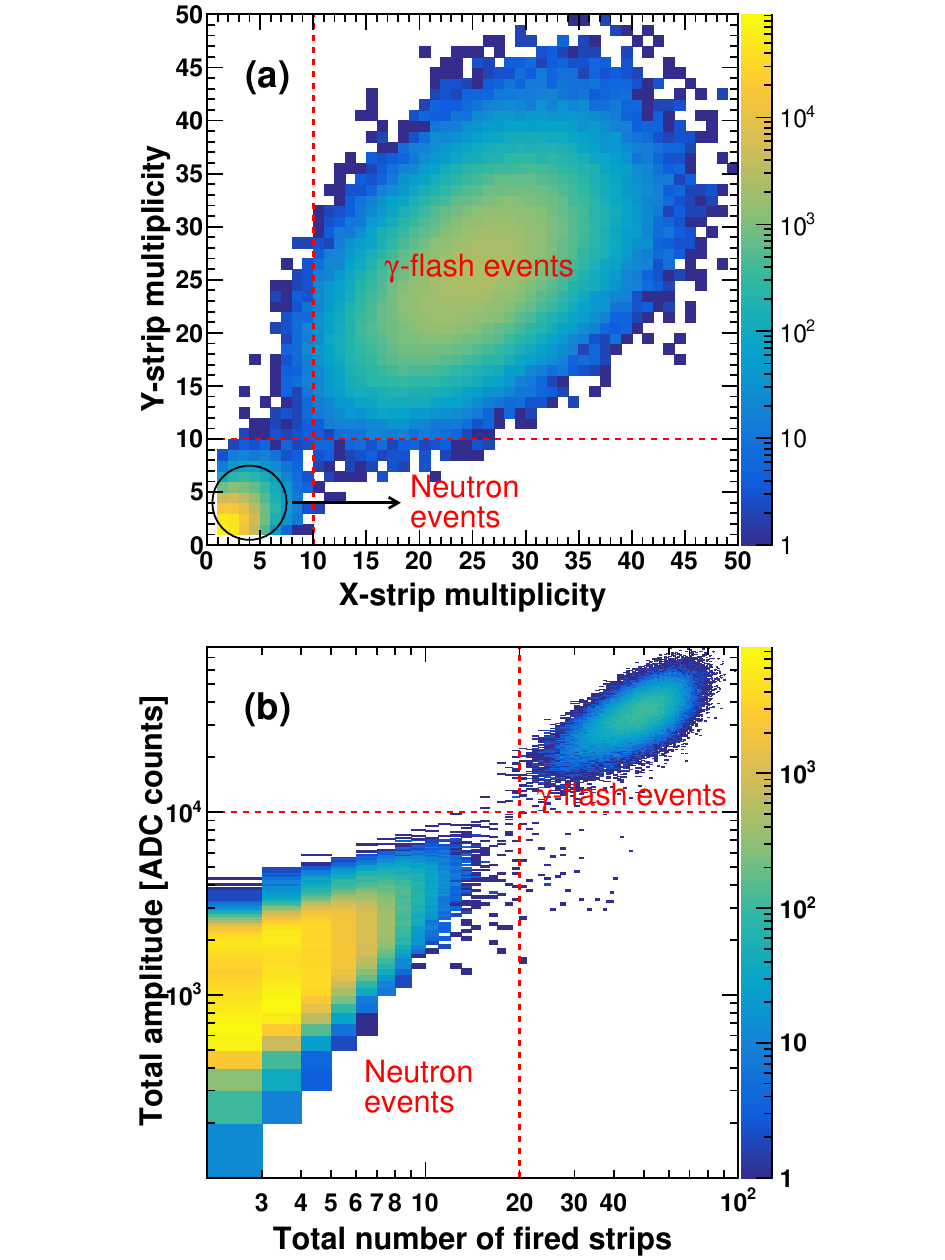}
  \caption{(a)~The distribution of the X- and Y-strip multiplicities.  (b)~Discrimination of $\gamma$-flash background events from neutron events by using the total number of fired strips and the sum of the amplitudes of all the fired strips.  The dashed lines represent the selection criteria for neutron events.  The neutron signal is well separated from the $\gamma$-flash background.}
  \label{fig:mult}
 \end{figure*}

The dominant background to the neutron events comes from the in-beam $\gamma$-flash produced by the impact of the \SI{1.6}{\GeV} proton beam on the tungsten target, which leads to the emission of a variety of particles other than neutrons, including an intense flux of $\gamma$ rays that propagates to the Back-n experimental halls through the beam pipe.  The use of a bending magnet, about \SI{20}{\m} away from the target, helps to remove the charged particles from the back-streaming neutron beam~\cite{bib:Zhang2018}.  Since the recording of the absolute time of the strip signal is not enabled in the readout electronics configuration for this measurement, the rejection of $\gamma$-flash background has to rely on the offline analysis.  As shown in Fig.~\ref{fig:mult}~(a), the multiplicities can be used to discriminate the neutron events and the $\gamma$-flash events.  Two well separated populations of events are clearly seen.  The low (high) multiplicity region, where the X- or Y-strip multiplicities are typically between 1 and 9 (between 10 and 40), corresponds to the neutron ($\gamma$-flash) events.  This characteristic of the multiplicity distribution is mainly due to that the $\gamma$-flash events simultaneously fire strips over the entire detector surface, resulting in a high multiplicity; while for a neutron event, only a few strips around the point of interaction of the neutron with the $^{10}\mathrm{B}$ layer give signals (called fired strips), resulting in a low multiplicity.  Together with the total signal amplitudes of all the fired strips, a clear separation between $\gamma$-flash background events and neutron events is obtained, as shown in Fig.~\ref{fig:mult}~(b).  

\subsubsection{Reconstruction of the beam spot distribution}
\begin{figure*}[p]
\centering
  \includegraphics[width=\textwidth]{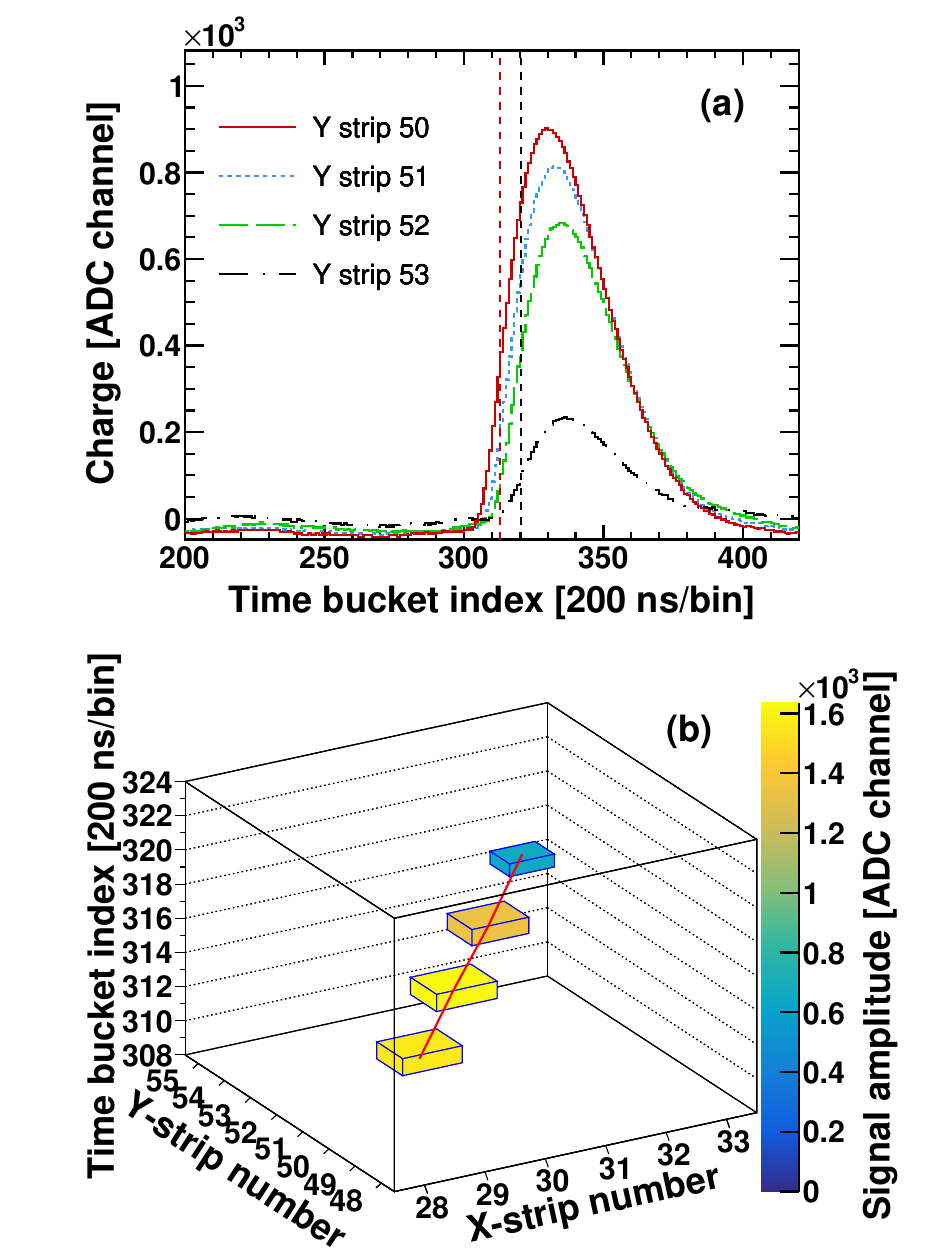}
  \caption{(a) The signal waveforms of four fired Y strips for a charged particle event.  The sampling frequency is \SI{5}{\MHz}.  The vertical red (black) dashed line represents the time of the earliest (latest) Y strip giving a signal, i.e. Y strip 50 (53).  (b) The charges collected by four consecutive X and Y strips along the charged particle trajectory in the gas, represented by the boxes.  The color code indicates the sum of X- and Y-strip signal amplitudes in units of [ADC channels].}
  \label{fig:track}
 \end{figure*}

After applying the event selection criteria described above, clean neutron events are obtained for the reconstruction of the beam profile.  The $\mu$TPC method is employed for the reconstruction, which uses the central positron of the latest strip as an estimation for the position of the neutron interaction in the converter.  The relative time information of each fired strip recorded in an event is determined by fitting the leading edge of the waveform to the Fermi--Dirac function.  The fitted time at half height is considered as the arrival time of the strip signal.  The fitting range is from 10\% to 90\% of the maximum height on the leading edge.  The strip with the earliest arrival time is taken as reference and the arrival time of adjacent strips is measured relative to it, as shown in Fig.~\ref{fig:track}.  A low sampling frequency (\SI{5}{\MHz}) was chosen to ensure a full waveform recording.
 
\begin{figure*}[p]
\centering
  \includegraphics[width=\textwidth]{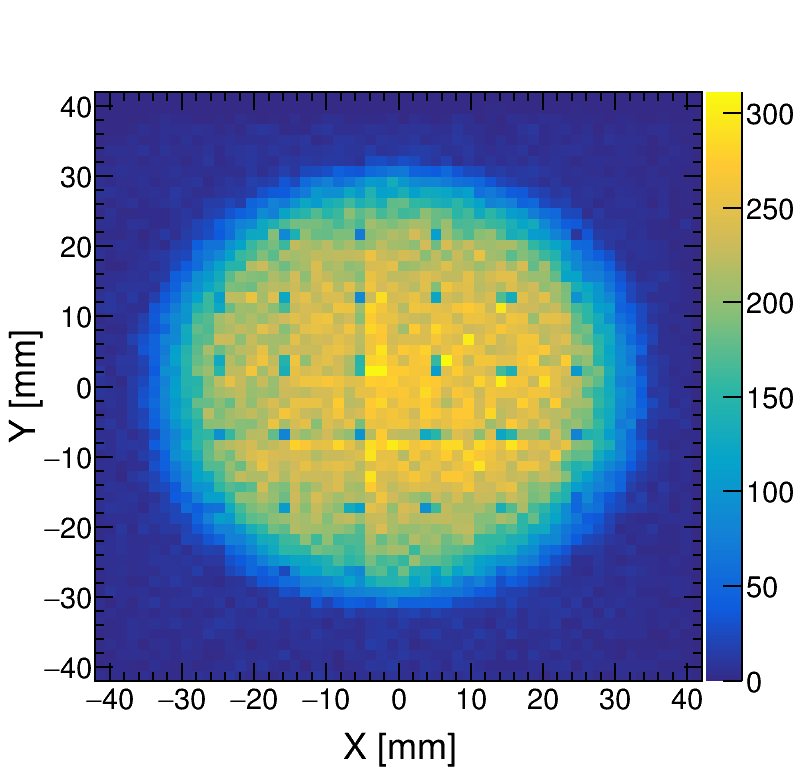}
  \caption{The reconstructed 2D profile of the Back-n white neutron beam at ES\#1.  The bin widths for the 2D histogram are \SI{1.5}{\mm} on both axes.  The shaded dots in the profile image correspond to the reconstructed positions of the pillars.}
  \label{fig:2Dprof}
 \end{figure*}

\begin{figure*}[p]
\centering
  \includegraphics[width=\textwidth]{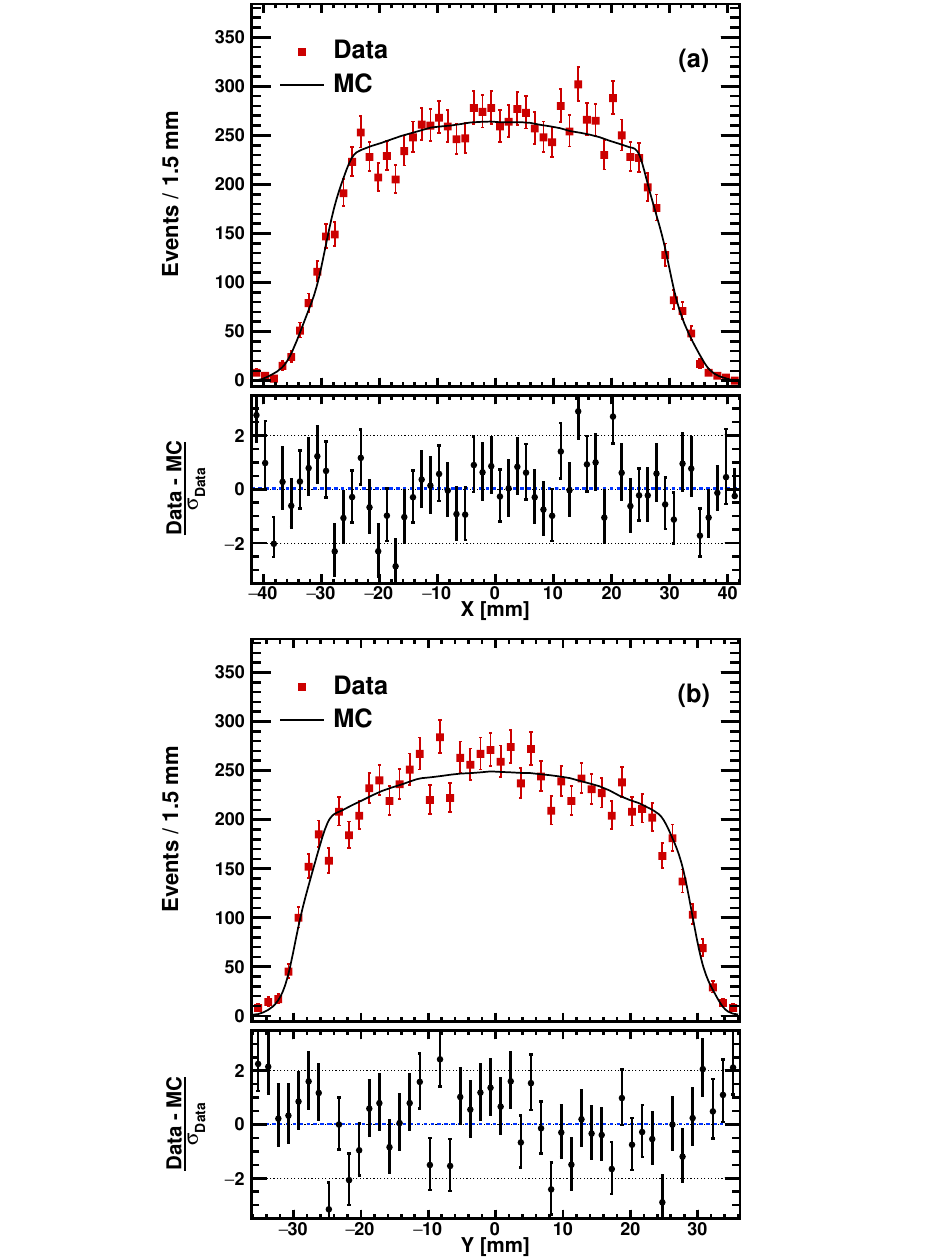}
   \caption{The upper panels show the 1D projections of the slices of the 2D profile corresponding to (a) the middle Y-strip onto the X-axis and (b) the middle X-strip onto the Y-axis, for both data (solid squares) and MC simulation (solid curves).  The expectation from MC simulation is normalized to data.  The lower panels contain the significance of the deviation between the observed data and the simulation prediction in each bin of the distribution, considering only the statistical fluctuations in data.}
  \label{fig:prof_xy0}
 \end{figure*}

Figure~\ref{fig:2Dprof} shows the reconstructed 2D profile of the Back-n white neutron beam at ES\#1.  The positions of the \SI{2}{\mm} diameter pillars are clearly visible.  However, these results cannot be directly used to estimate the detector spatial resolution.  Placing masks with different shapes like rectangular or circular holes in front of the detector will be part of the future experiment with neutron beams, so that the spatial resolution can be directly estimated.  Figures~\ref{fig:prof_xy0}~(a) and (b) show the 1D projections of the slices of the 2D profile corresponding to the middle Y- and X-strips onto the X- and Y-axis, respectively, for both data and MC simulation.  As can be seen, the measurement and the simulation prediction agree within uncertainties.  The FWHM of this projection on the X-axis (Y-axis) is \SI{\sim58}{\mm} (\SI{\sim54}{\mm}) according to the simulation.  The reconstructed 2D Back-n beam profile is not circular because the 2D proton beam profile at the spallation target does not have circular symmetry.  The widths of the horizontal and vertical profiles of the proton beam are \SI{\sim88}{\mm} and \SI{\sim33}{mm} (FWHM), respectively.
 

\section{Conclusions and outlook}

Two variants of the 2D readout Micromegas-based neutron detector have been fabricated and characterized with a ${}^{55}\mathrm{Fe}$ X-ray source to obtain the relative electron transparency, the gain and the gain uniformity, and the energy resolution.  The detectors have been tested with respect to the neutron beam profile reconstruction capability using an ${}^{241}\mathrm{Am}$ $\alpha$ source, an Am-Be neutron source, and the CSNS Back-n neutron beam.  In the test with the Am-Be source (Back-n), a ${}^{6}\mathrm{Li}$ (${}^{10}\mathrm{B}$) converter was used with the copper-anode (resistive-anode) Micromegas detector.  The results of the tests show that the detector performance meets the basic requirements for neutron beam profile measurements.  Then, the neutron beam profile of the CSNS Back-n was measured using the copper-anode Micromegas detector with ${}^{10}\mathrm{B}$ converter and a 2D neutron beam spot distribution at the Back-n Endstation-1 was obtained.  The good agreement between the simulations and experimental data confirms the reliability of this measurement.

The recording of the absolute time of the strip signal will be added in the front-end electronics configuration as part of the further development, such that (a) an online rejection of the $\gamma$-flash background and (b) measurement of the neutron TOF or energy spectrum will become possible.  Other improvements will also be carried out, such as (a) the increase of the length of the drift gap in order to obtain a higher strip multiplicity for a more precise determination of the neutron interaction point and (b) the reduction of the size of the pillars in order to lessen the neutron detection inefficiency caused by the pillars.  Recently, the fabrication technique has been improved, reducing the diameter of the pillars to \SI{0.5}{\mm}.

\section{Acknowledgements}
This work was supported by the National Key R\&D Program of China [grant number 2016YFA0401600]; the National Natural Science Foundation of China [grant numbers 11605218, 11605197]; the Fundamental Research Funds for the Central Universities; and the State Key Laboratory of Particle Detection and Electronics, China [grant numbers SKLPDE-ZZ-201801, SKLPDE-ZZ-201818].  This work was partially carried out at USTC Center for Micro- and Nanoscale Research and Fabrication, and we thank Yu Wei for his help on the nanofabrication steps for germanium coating.


\bibliography{ylmmpaper}

\newpage

\end{document}